\documentclass[useAMS,usenatbib]{mn2e}

\usepackage{graphicx}
\usepackage{times}
\usepackage{natbib}

\begin{document}
 
\title[Turbulent fields in simulated ICM: scaling laws]{Turbulent Velocity Fields in SPH--simulated 
Galaxy Clusters: Scaling Laws for the Turbulent Energy}
\author[F. Vazza, G. Tormen, R. Cassano, G. Brunetti, K. Dolag]{F. Vazza$^{1,2,3}$%
\thanks{%
 E-mail: vazza@pd.astro.it}, G. Tormen$^{1}$, R. Cassano$^{2,3}$, G. Brunetti$^{3}$, K. Dolag$^{1,4}$ 
\\
$^{1}$Dipartimento di Astronomia, Universit\'a di Padova, vicolo
dell'Osservatorio 2, 35122 Padova, Italy \\
$^{2}$ Dipartimento di Astronomia, Universit\'a di Bologna, via Ranzani
1,I-40127 Bologna, Italy\\
$^{3}$ INAF/Istituto di Radioastronomia, via Gobetti 101, I-40129
Bologna, Italy\\
$^{4}$ Max-Planck-Institut f\"ur Astrophysik, Garching, Germany}
\date{Accepted 2006 February 28; Received 2006 February 23; in original form 2006 February 8}
\maketitle

\begin{abstract}
We present a study of the turbulent velocity fields in the Intra
Cluster Medium of a sample of 21 galaxy clusters simulated by the
SPH--code Gadget2, using a new numerical scheme where the artificial viscosity
is suppressed outside shocks. The turbulent motions in
the ICM of our simulated clusters are detected with a novel method devised to better
disentangle laminar bulk motions from chaotic ones. 
We focus on the scaling law between
the turbulent energy content of the gas particles and the total mass, and find that
the energy in the form of turbulence scales approximatively with the thermal 
energy of clusters. 
We follow the evolution with time of the scaling laws and discuss the physical
origin of the observed trends. The simulated data are in agreement with independent
semi--analytical calculations, and the combination between the two methods allows 
to constrain the scaling law over more than two decades in cluster mass.

\end{abstract}

\label{firstpage}

 
\begin{figure*}
\includegraphics[width=0.30\textwidth]{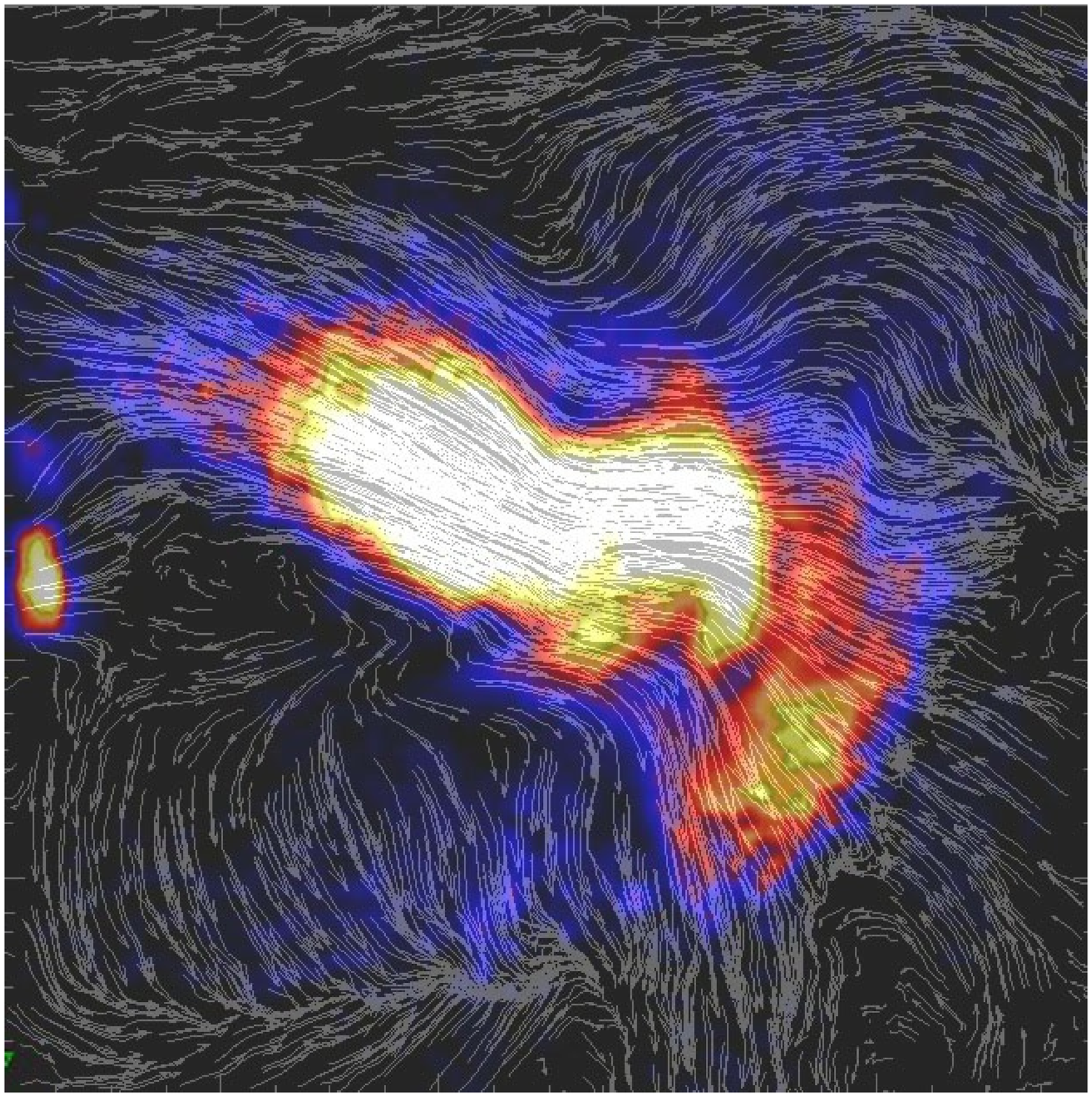}
\includegraphics[width=0.30\textwidth]{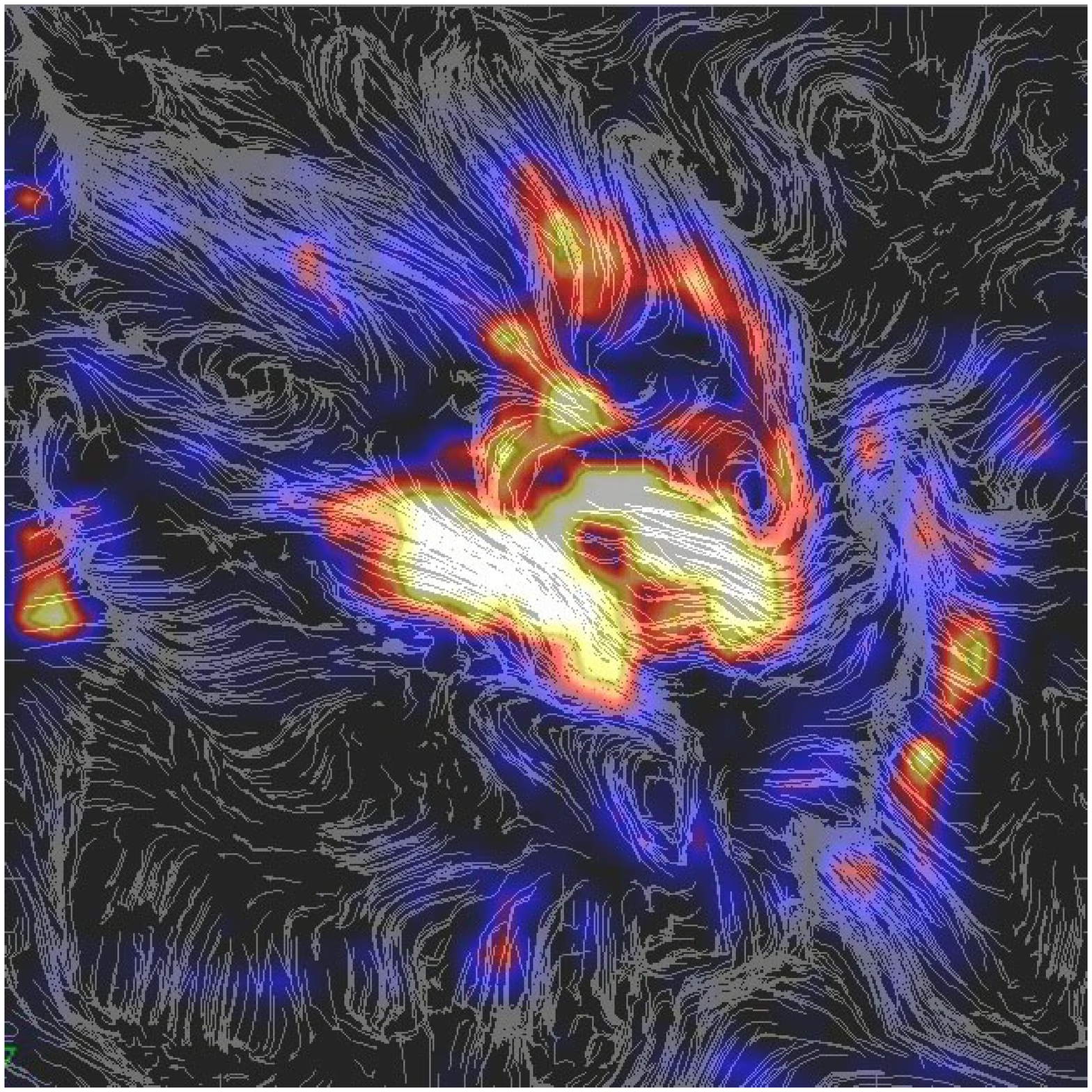}
\includegraphics[width=0.30\textwidth]{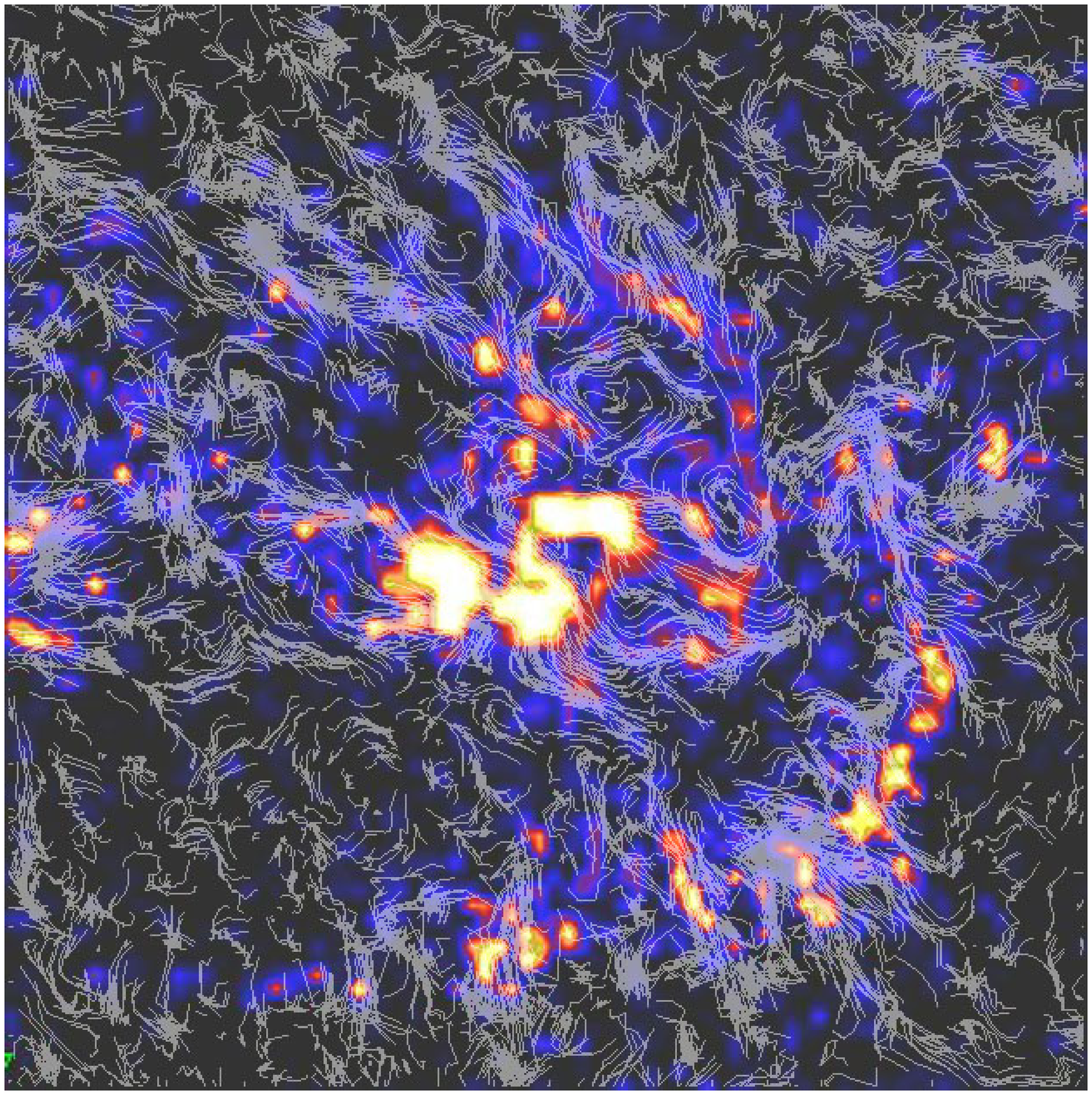}
\caption{2--dimensional maps for the central $\sim (1.3 \rm{Mpc})^{2}$
region of the most massive cluster of our sample. A subcluster ten
times less massive is being accreted (entering from the left). The
streamlines on the {\it left} depict the velocity defined with respect
to the center of mass velocity; on the {\it right} depict the local
velocity residuals for two different resolutions of the algorithm:
$l\sim 32 \rm{kpc}$ ({\it center}) and $l\sim 16 \rm{kpc}$({\it right}).
The overlayed colour maps show the kinetic energy of gas 
particles for each of the velocity fields.}
\label{fig:maps}
\end{figure*}

 \section{Introduction}
\label{sec:intro}

A good deal of evidence from observational and numerical works is
 suggesting that a non negligible budget of turbulent
motions is stored in the Intra Cluster Medium (ICM) of galaxy
clusters. The observational evidence from the gas pressure map of the Coma
cluster (Schuecker et al. 2004), the lack of resonant scatterings in
Perseus (Churazov et al. 2004), and the many indications of
non--thermal emissions possibily due to acceleration of particles into
a turbulent magnetized medium (e.g. Brunetti et al.2004 and references
therein for a review),
are pointing towards the relevance of chaotic motions within the ICM. 
Moreover, Eulerian numerical simulations of merging clusters (e.g.,
Bryan \& Norman 1998; Roettiger, Stone \& Burns 1999; Ricker \& Sarazin 2001) 
or of accretions of
less massive structures as dwarf galaxies (e.g. Mayer et al. 2005) have
provided good representations of the way in which turbulence may be
injected into the ICM.  Also theoretical calculations expect a
non--negligible amount of
turbulent motions in galaxy clusters (e.g. Cassano \& Brunetti 2005, 
C\&B05), possibly with a relevant amplification of 
seed magnetic fields through kinetic
dynamo--effects (e.g. Subramanian, Shukurov \& Haugen 2005; En\ss lin
 \& Vogt 2005; Schekochihin \& Cowley 2006). Even the claim against
 turbulence in the ICM of some observed galaxy clusters as Perseus
 (e.g. Fabian et 
al. 2003) actually rule out only scenarios of very strong turbulence, i.e. with
an energy density in turbulence not far from the thermal support.

Unluckily, due to the abrupt breaking of the main instrument onboard
of the Astro-E2 mission, the direct detection of turbulent fields
through the broadening of iron lines profile (e.g. Inogamov \& Sunyaev
2003) has to be postponed to the future.  From the numerical
viewpoint, Smoothed--Particles--Hydrodynamics (SPH) represents a
unique tool to investigate the physics of both the collisional (gas)
and non collisional (dark matter - DM) mass components of galaxy
clusters; moreover, the high dynamic coverage of SPH permits to study
a large interval of cluster sizes, and allows one to follow the
evolution of the most interesting features with cosmic time. In this
Letter we present the first results from a quest for turbulent
velocity fields in a sample of well-tested SPH simulated galaxy
clusters (Sec. \ref{sec:simulations}); so far, this is the first
time that a fully cosmological and highly resolved set of simulations
is employed to search and characterize ICM turbulence.  To
this end we developed a novel approach to the problem of disentangling
bulk motions of gas particles from chaotic ones
(Sec. \ref{sec:turbulence}); in Dolag et al. 2005 (hereafter Do05) we
 claimed that such method is more
efficient than only removing the cluster bulk velocity as usually done
in literature (e.g. Bryan \& Norman 1998), allowing one to focus 
only on the most chaotic part of the
ICM flow.  
In this Letter we focus on the scaling
relation between the thermal energy content of simulated
galaxy clusters and their kinetic energy in the form of
turbulent motions (Sec. \ref{sec:laws}) and finally 
compare our findings with the expectations from semi--analytical 
calculations (Sec. \ref{sec:semianalytical}).

\section{The sample of simulated clusters.}
\label{sec:simulations}
The detailed information about our cluster sample can be found in Do05. 
It consists of 9 resimulations with 21 galaxy
clusters and groups simulated with the tree N-body--SPH code Gadget2 (Springel
2005), performed several times with different physical processes. The cluster regions were extracted 
from a dark matter only
simulation with box of $479\,h^{-1}$Mpc on a side and in the context of
a  $\Lambda$CDM model with
$\Omega_0=0.3$, $h=0.7$, $\sigma_8=0.9$ and $\Omega_{\rm
b}=0.04$ (Yoshida, Sheth \& Diaferio 2001). Adopting the `Zoomed Initial Conditions' technique 
(Tormen, Bouchet \& White 1997) the regions 
were re--simulated in order to achieve
higher mass and force resolution.  Particle mass for the resimulations
is $m_{\mathrm{DM}}=1.13\times 10^{9}\,h^{-1}M_{\odot}$ and
$m_{\mathrm{gas}}=1.7\times 10^{8}\,h^{-1}M_{\odot}$; the
gravitational softening adopted is $h_{i}\simeq 5\,h^{-1}$kpc (Plummer
equivalent between $z=5$ and $z=0$ and kept fixed in comoving units at 
higher redshits; this also marks the minimum value for smoothing the gas
particles within the SPH algorithm). 

Here we used all the 21 haloes having virial masses in the range
 $M_{\rm{vir}}=5.3\cdot 10^{13} - 2.3 \cdot 10^{15}M_{\odot} \cdot h^{-1}$ found
in the set of simulations. They are resolved by
$8\cdot 10^{4}$ to $4\cdot 10^{6}$ gas and DM particles respectively.
Even if the whole
sample allows one to study the role of plasma conductivity, cooling,
star formation and feedback processes, we restricted so far to a
non-radiative SPH subset where an improved recipe for the numerical
viscosity of gas--particles is used.
This recipe, which follows an idea
of Monaghan \& Morris (1997) manages to reduce the amount of
numerical, and therefore un--phyisical dissipation of chaotic motions, ensuring at the same
time a good treatment of shocked features; its use is mandatory for any study
of the chaotic part of the ICM dynamics, since otherwise the standard SPH 
scheme greatly suppresses any chaotic motion at the smallest scales, 
even in absence of shocks. 
For a more detailed discussion about this method we
address the reader to Do05.

\section{Detection of turbulent motions.}
\label{sec:turbulence}
 In order to characterize turbulent velocity fields in a fluid medium,
the crucial point is to extract a pattern of velocity
fluctuations from a complex distribution of velocities. The most used
approach in the literature (e.g. Norman \& Bryan 1998; Sunyaev, Norman 
\& Bryan 2003) is simply to use individual velocities of the gas particles
after subtracting of the mean velocity, computed within a fixed volume.
Although this method has been widely employed in many previous works, 
which indeed found a remarkable level of
turbulence within simulated ICM, it can turn out to be highly 
misleading in the 
case of substructure crossing the cluster volume. As shown in the left
panel of Fig. \ref{fig:maps}, the motion of a subcluster usually tracks
laminar velocity patterns which can differ from the mean virial velocity.
This provides a spurious contribution to
the estimated turbulent energy, whereas
the only correct contribution to consider is the chaotic velocity field
at the interface layer with the resident ICM of the primary cluster and
along the tail of the subcluster.
In order to improve on this we conceived an
algorithm which disentangles the chaotic part of the flow in a more
effective way. We defined a mean local velocity field $\bf{v}_{\rm{cell}}$
 in a cell by
interpolating the velocity and density of each gas particle onto a
regular mesh using a {\it Triangular Shape Cloud} (TSC) window
function. The dependence of the mesh--spacing to the final results is
discussed in section \ref{sec:laws}. We then evaluated the local
velocity fluctuations of each gas particle by:
\begin{equation}
\bf{\delta v}_{\rm{i}} \simeq \bf {v}_{\rm{i}} - \bf{v}_{\rm{cell}},
\label{sigmai}
\end{equation}
where $\bf{v}_{\rm{i}}$ is the 3--dimensional particle velocity,
$\bf{v}_{\rm{cell}}$ is the local mean velocity field of the cell where
each particle falls. This field obviously varies depending on the
choice of grid size: the subtraction of a local velocity field is
expected to progressively filter out the contribution from laminar
bulk--flows produced by the gravitational infall (which is certainly
relevant, for any redshift and all distances from the cluster center, 
e.g. Tormen, Moscardini \& Yosida 2004). Figure \ref{fig:maps}
gives an example of the velocity fields obtained by subtracting the mean
bulk velocity ({\it left panel}) and after subtracting the local mean velocity
found by our new algorithm for two representative resolutions
of the TSC algorithm ({\it middle and right panel}).
The laminar flow at the largest scales is filtered out by the
TSC--kernel, while the most chaotic and swirling flows are
highlighted, even though some of the biggest vortices might also have
been filtered out.
We stress here that this method is just a
zero--order approximation: an accurate spectral analysis for turbulent
dynamics within the simulated ICM will require further improvements,
since the ICM is likely characterized by multiple energy injections (from
mergers and accretions) driven simultaneously from different scales.

\begin{figure}
\includegraphics[width=0.23\textwidth]{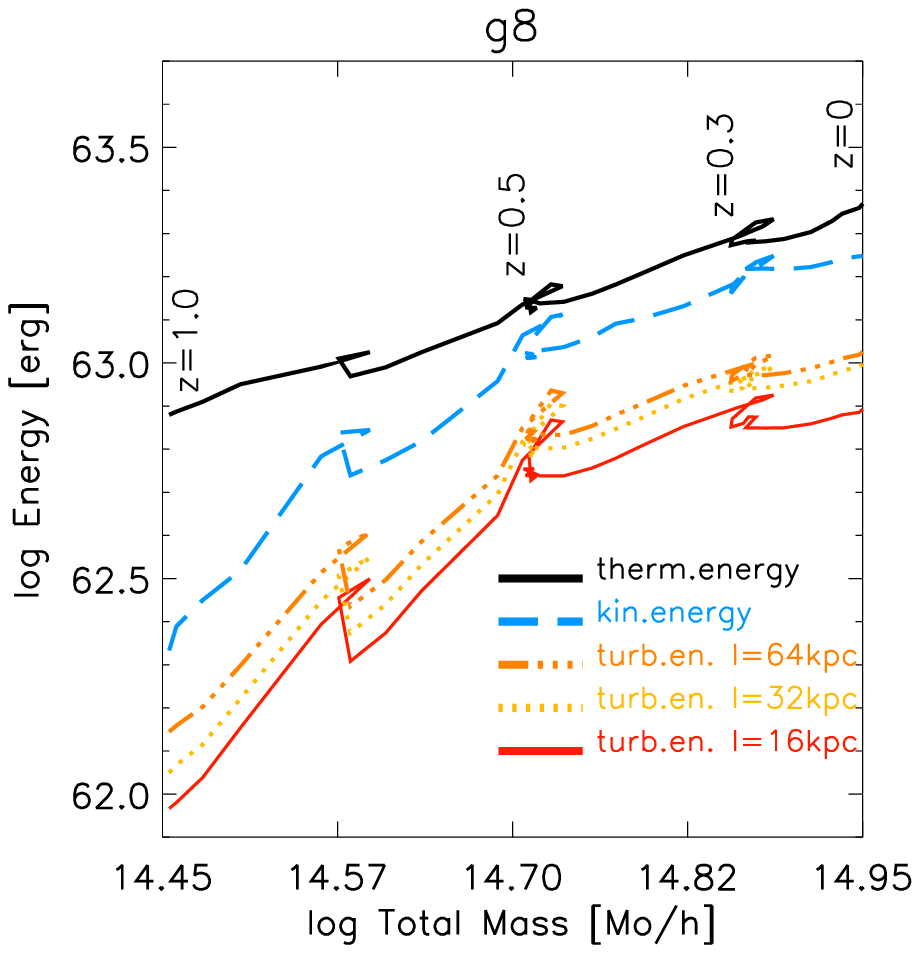}
\includegraphics[width=0.23\textwidth]{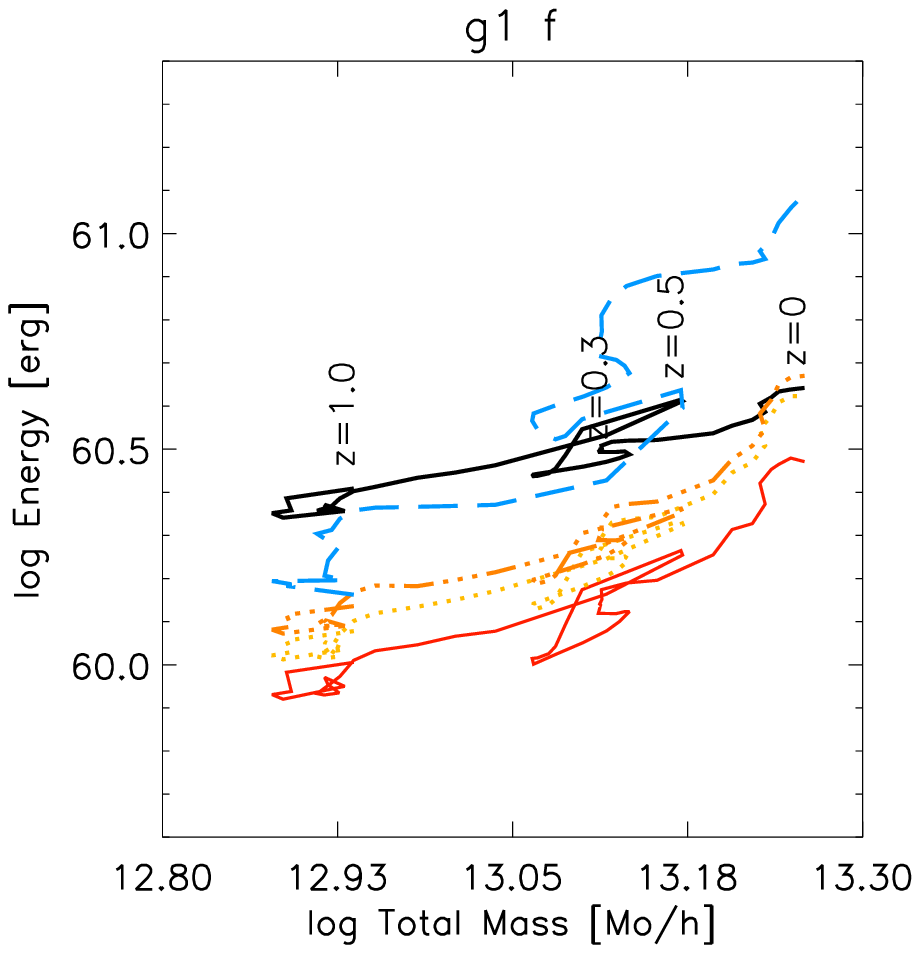}
\includegraphics[width=0.23\textwidth]{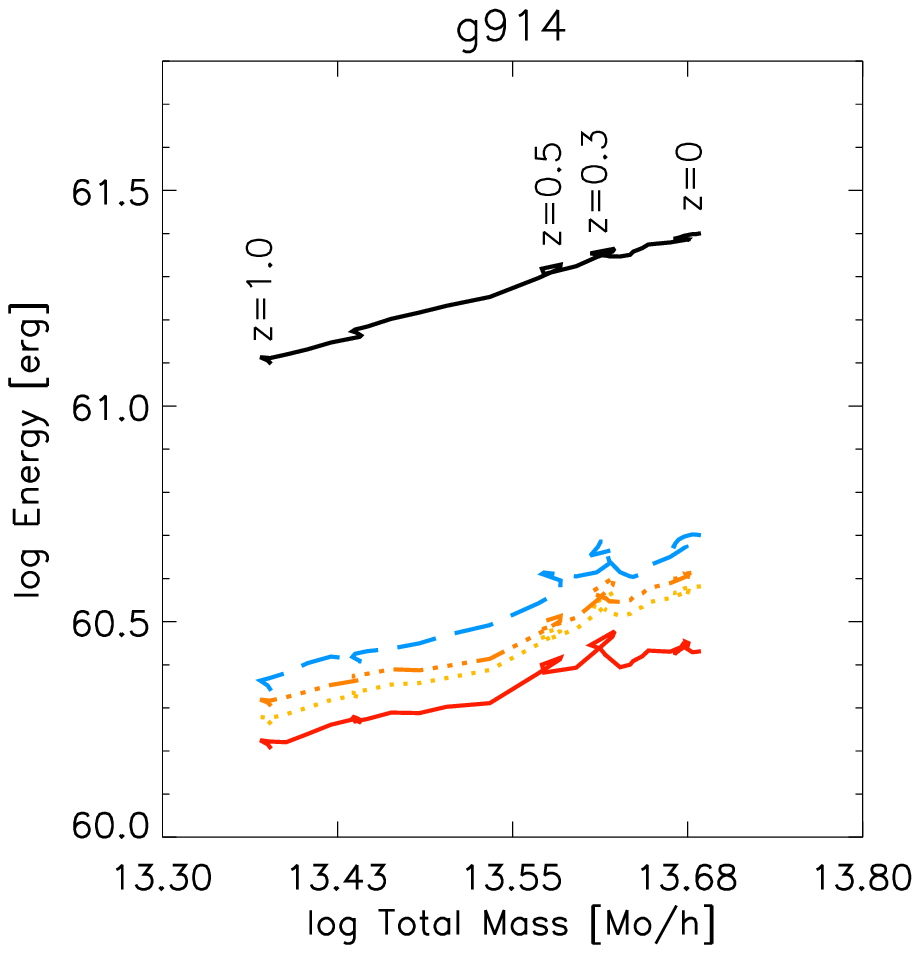}
\includegraphics[width=0.23\textwidth]{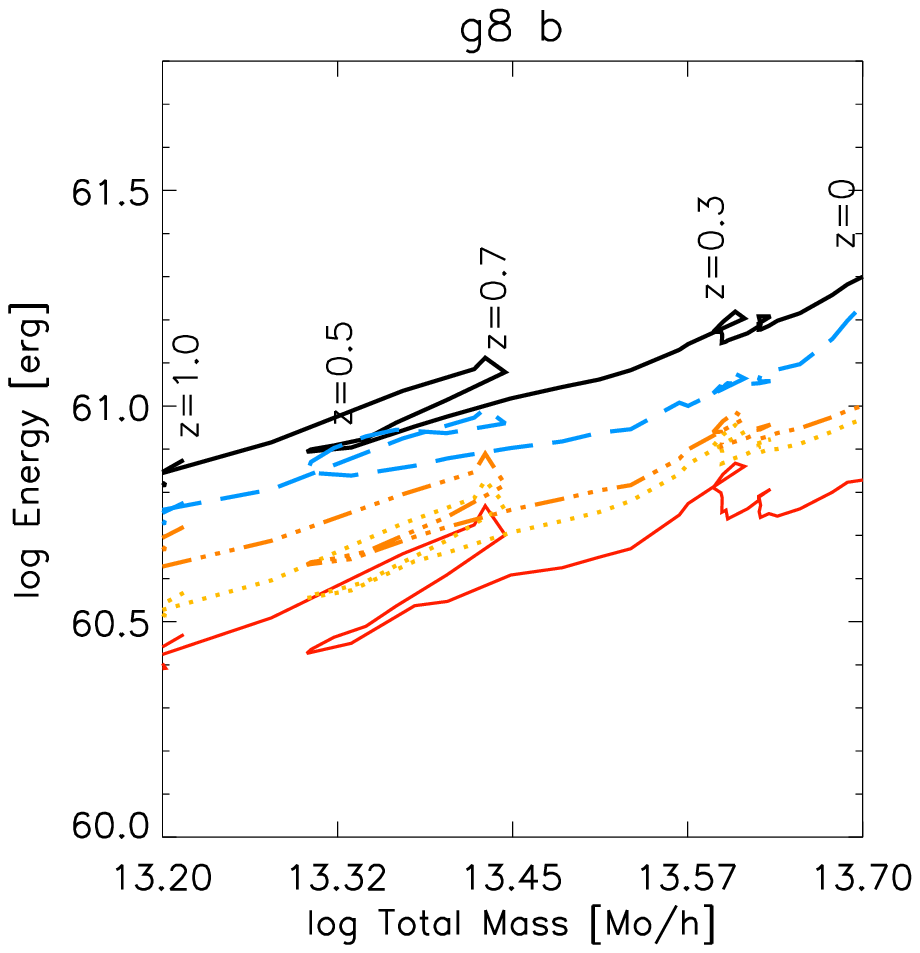}
\caption{Individual paths for fours clusters of the sample, in the
$\log \rm{Energy}-\log \rm{M_{\rm{tot}}}$ plane. The upper two panels show
the evolution of the most and of the less massive cluster within our 
catalogue, whereas the lower two panels show the evolution of two clusters with
a nearly equal final mass ($\simeq 5\times 10^{13}M_{\odot}h^{-1}$), but very different
``relaxation'' state: left panel is for the ``relaxed'' (i.e. $\xi < 0.5$ at z=0) cluster g914 while 
the right one is for the  ``perturbed'' one, g8 b ($\xi \geq 0.5$ at z=0).}
\label{fig:single}
\end{figure}

\section{Scaling laws for Turbulent Kinetic Energy}
\label{sec:laws}

 The main goal of this section is to 
investigate the scaling laws between the mass (gas plus
dark matter particles) of clusters/groups, $M_{\rm{tot}}$,
and the thermal, kinetic and turbulent energy of the ICM.

Due to computational limitations we so far restricted our analysis to a cubic
region, centered onto the center of the cluster, of equivalent volume
$V_{\rm{box}}=(R_{\rm{vir}})^{3}$. This ensures that we consider in any case a
number of gas particles ranging from several thousands to nearly one million. 
After the velocity decomposition is performed (section 
\ref{sec:turbulence}), we evaluate the turbulent energy
content as:
\begin{equation}
E_{\rm{TUR}}=\frac{1}{2}m_{\rm{gas}}\sum_{\rm{BOX}}\delta v_{i}^{2}
\label{eq:general}
\end{equation}
where the sum is done over the module of the velocity fluctuation, 
$ \delta v_{i}$, of the gas particles.
This calculation was repeated at three different resolutions of the
TSC--kernel used to define the local mean velocity field: $ l=$16, 32
and 64 kpc.  As discussed in Do05 a reasonable subtraction of the laminar
 pattern is obtained with a mesh spacing around $\sim 30 $kpc (corresponding
to a TSC equivalent length three times larger)\footnote{Note that
this filtering width is always larger than the smoothing
length of gas particles within the volume region we consider, for each
cluster and each redshift of observation. 
An adaptive sampling scheme would be expecially 
mandatory to catch turbulence in
the cluster outskirts, due to the decrease of particle density.}.

The total kinetic and thermal energies were
evaluated as:
\begin{equation}
E_{\rm{TH}}=\frac{3}{2}m_{\rm{gas}}\sum_{\rm{BOX}}\frac{f_{e}k_{B}T_{i}}{\mu
m_{p}},
\end{equation}
where $k_{B}$ is Boltzmann's constant, $T_{i}$ the gas particle
temperature, $\mu=0.59$ the mean molecolar weight in AMU, $m_{p}$
the proton mass and $f_{e}=0.58$ the fraction of free electrons per
molecule, assuming a primordial mixture of $x_{H}=0.76$, and
\begin{equation}
E_{\rm{K}}=\frac{1}{2}m_{\rm{gas}}\sum_{\rm{BOX}}v_{i}^{2},
\end{equation}
where the module of velocity, $v_{i}$, has been reduced to the center of mass
velocity frame (as in Norman \& Bryan 1998).

In figure \ref{fig:single} we report the time evolution
of four representative clusters in our sample
in the $E_{\rm{TUR,TH,K}}$--$M_{\rm{tot}}$ plane.
The most ``relaxed'' structures (as the cluster g914, 
bottom left panel)
present a fairly smooth evolution, whereas ``perturbed''
structures (as g8b and g1f, right panels) show
a more complex evolution with episodic jumps
in turbulent and kinetic energies, and
have a high ratio, $\xi$, between kinetic (and turbulent)
and total (thermal plus kinetic) energy. 
This reflects the significant difference
in the ratio between the kinetic and the potential energy of these clusters
(e.g. Tormen et al.~1997), which is higher for the perturbed ones.

Since our cluster sample is extracted from re--simulations
centered on 9 massive and fairly isolated clusters, smaller
systems generally correspond to structures about to be accreted 
by larger ones.  As such, small systems are often perturbed, and this 
introduces a bias in the dynamical properties of the cluster population.
This bias can however be alleviated by restricting our analysis only to
the most ```relaxed'' objects in our sample, as we will see below.

\noindent
In general, we find the following power law scaling between cluster 
energy (thermal, kinetic or turbulent) and cluster mass:
\begin{equation}
E_{j} \sim A_j (\frac{M_{\rm{tot}}}{10^{15}M_{\odot}h^{-1}})^{D_{j}},
\label{eq:plan}
\end{equation}

\noindent
with $j=TH$, $K$, $TUR$, and where $A_j$ and $D_j$ are the zeroth point and the
slope of the correlations, respectively.

\noindent
We find that the scaling of thermal energy with mass is always consistent
with that expected in the virial case, $D_{\rm{TH}} \sim 5/3$,
while the values of $D_{K}$ and $D_{TUR}$ slightly depend on 
the number of "perturbed" small systems included in the analysis.
With all system included, the slope of the scaling between turbulent 
energy and cluster mass is flatter than that between thermal energy 
and mass by $\sim 0.2$. As we remove more and more small perturbed systems,
the turbulent slope steepens toward the thermal value.
If we define the ratio $\xi$ between turbulent and thermal energy for
each system and redshift, we find that the flattening of the turbulent 
scaling with respect to the thermal scaling is statistically significant 
only if objects with $\xi \geq 0.5$ (nine at z=0) are included.

\begin{figure}
\includegraphics[width=0.48\textwidth]{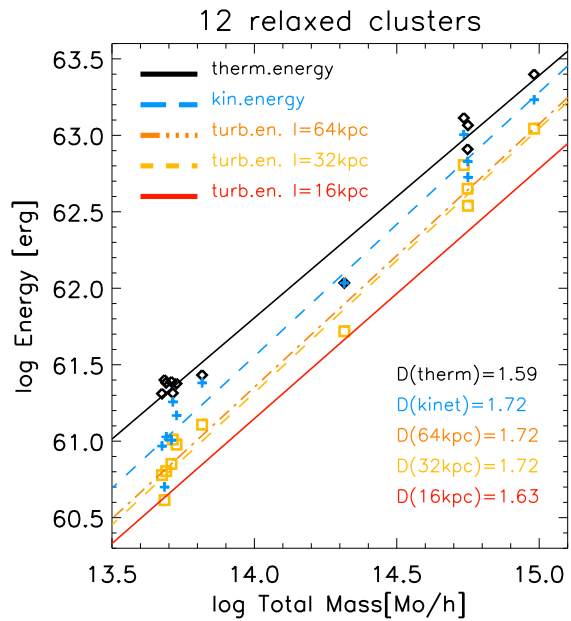}
\caption{Scaling laws at redshift $z=0$ for the 12 most relaxed clusters 
($\xi < 0.5$); the values of the slopes
for the different relations are reported in the panel. For the sake of
displaying,  only the datapoints of the $l=32kpc$ grid turbulence are drawn.}
\label{fig:z}
\end{figure}

The slopes of the scalings thus obtained are stable and do not depend
on the value of the mesh--spacing, $l$, adopted for the subtraction of the laminar
motions (Fig.~\ref{fig:z}), as the maximum difference does not exceed 
$\Delta D_{\rm TUR} \sim 0.1$ at any redshift.

In addition, the dependence of the turbulent energy on the adopted 
mesh--spacing confirms the behaviour already presented in Figure 4 of Do05,
with $E_{\rm{TUR}} \propto l^{1/2}$.

\begin{table}
\caption{Values for the slopes of the kinetic and turbulent 
scaling laws at zero redshift, for the 
whole sample of data and the ``relaxed'' subsample, with $1\sigma$ errors.}
\begin{tabular}{|c|c|c|c|c}
\hline
{\bf $ l $} & {\bf $D_{\rm TUR}$ (all)} & {\bf $D_{\rm TUR}$ (relax)} & {\bf $D_{\rm K}$ (all)} & {\bf $D_{\rm K}$ (relax)}\\
{\bf 16 kpc} & 1.43 $\pm$ 0.06 & 1.63 $\pm$ 0.04 & 1.38 $\pm$ 0.04 & 1.72 $\pm$ 0.03\\
{\bf 32 kpc} & 1.49 $\pm$ 0.04 & 1.72 $\pm$ 0.01 & 1.38 $\pm$ 0.04 & 1.72 $\pm$ 0.03\\
{\bf 64 kpc} & 1.49 $\pm$ 0.03 & 1.73 $\pm$ 0.05 & 1.38 $\pm$ 0.04 & 1.72 $\pm$ 0.03\\
\hline
\end{tabular}
\label{tab:tab1}
\end{table}

\begin{figure} 
\includegraphics[width=0.24\textwidth]{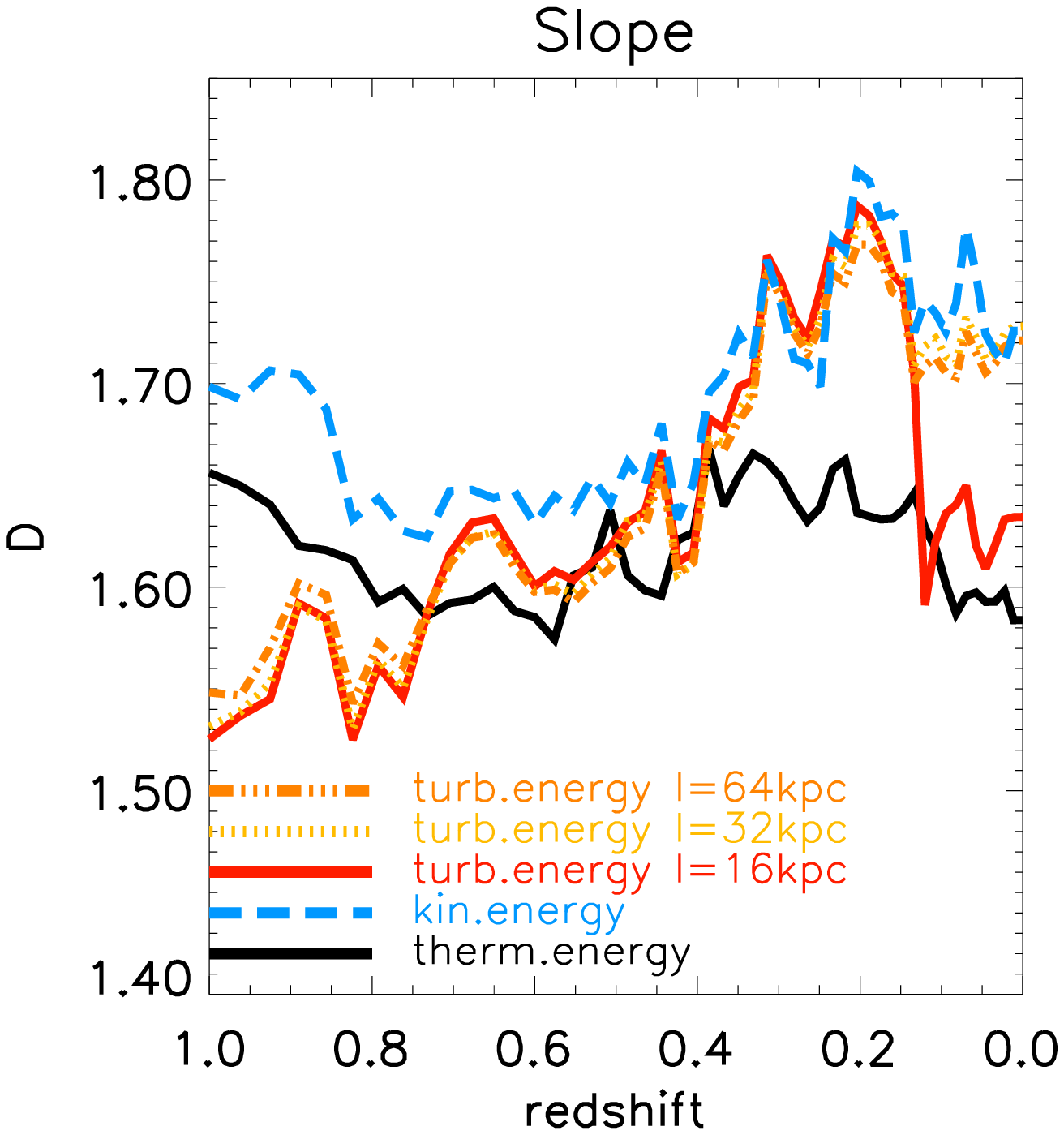}
\includegraphics[width=0.24\textwidth]{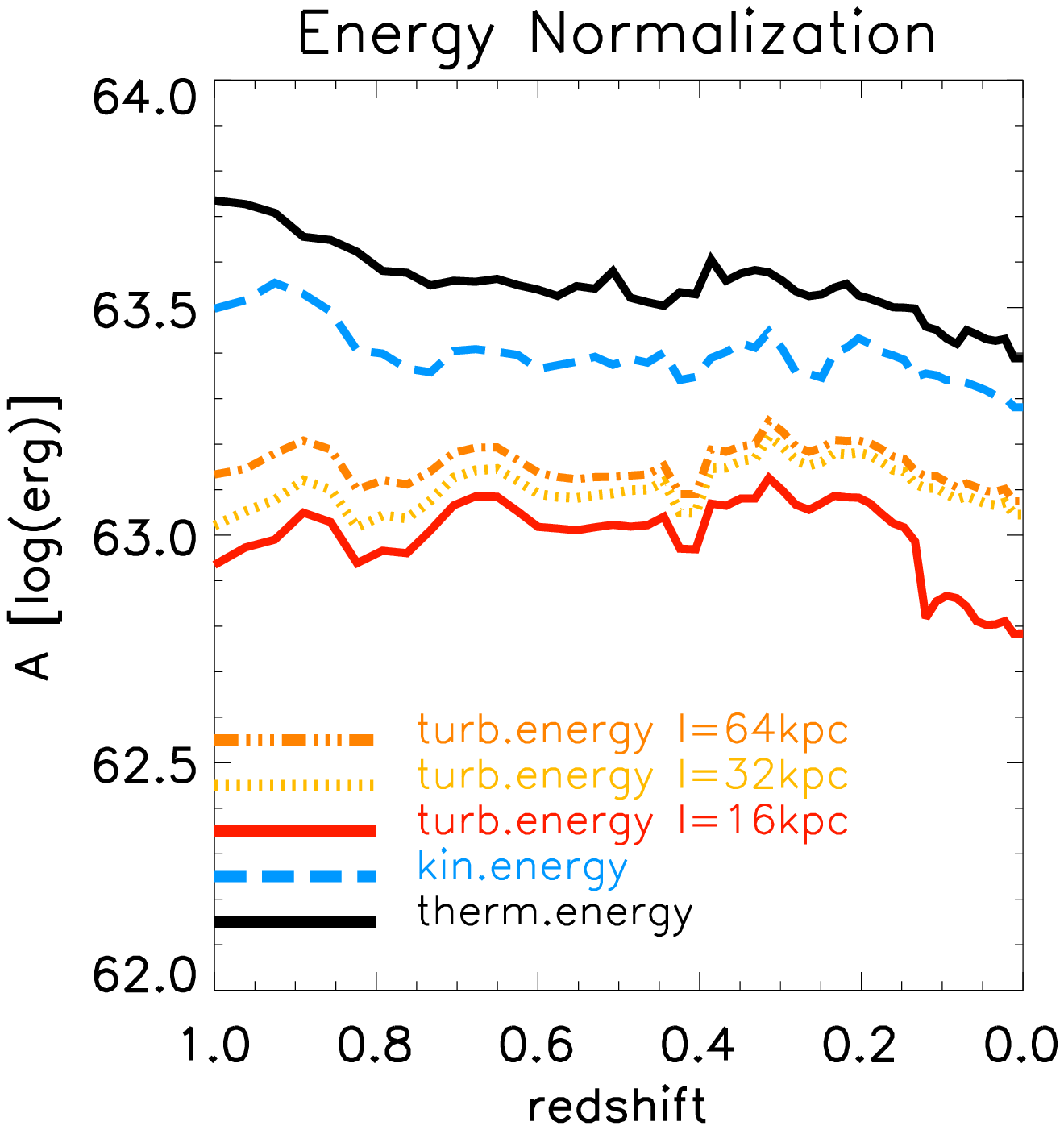}
\caption{Redshift evolution of slope({\it left}) and zero point ({\it right}) 
of the scaling law Eq.(\ref{eq:plan}), for the sample of objects with
$\xi < 0.5$.}
\label{fig:rel}
\end{figure}

Finally, Figure \ref{fig:rel} shows the redshift evolution 
of the slopes, $D_j$, and of the zero points, $A_j$, 
of the five correlations (Eq.~\ref{eq:plan}).
It is clear that the slopes are relatively constant with
redshift; this does not change significantly, unless
very ``perturbed'' groups
with $\xi \geq 0.5$ (at each z) are considered in the
analysis. In this last case a sistematic
flattening ($\Delta D_{\rm K,TUR} \sim 0.2$) of the scaling of the kinetic 
and turbulent energies  with cluster mass at low redshift is found: 
this is caused by the interactions between objects, which makes the 
smaller systems more and more perturbed as time proceeds.

\section{Comparison with semi--analytical results}
\label{sec:semianalytical}

  In the previous section we reported on the 
scaling between the turbulent energy and the
thermal (and kinetic) energy as measured in simulated clusters,
without motivating their physical origin.
Cluster mergers are likely to be responsible of
most of the injection of turbulent velocity fields in the ICM.
Simulations have indeed shown that the passage of a massive sub--clump through
a galaxy cluster can trigger turbulent velocity fields in the
ICM (Roettiger et al.1999; Ricker \& Sarazin 2001; Do05).

A simple method to follow the injection of merger--turbulence
during cluster life is given by semi--analytical calculations.
C\&B05 used merger trees to
follow the merger history of a synthetic population
of galaxy clusters (using the Press \& Schechter 1974 model)
and calculated the energy of the turbulence injected in the ICM 
during the mergers experienced by each cluster.
In these calculations 
turbulence is injected in the cluster volume swept by the sub-clusters,
which is bound by the effect of the ram pressure stripping,
and the turbulent energy is calculated as a fraction of the 
${\rm PdV}$ work done by the sub-clusters infalling onto the main cluster.

Although simplified, this semi--analytical approach 
allows a simple and physical understanding of the scaling
laws reported in the previous Section.
Indeed, since the infalling sub-clusters are driven by the gravitational
potential, the velocity of the infall should be 
$\sim 1.5-2$ times the sound speed
of the main cluster; consequently, the energy density of the turbulence injected during
the cluster--crossing should be proportional to the thermal energy density 
of the main cluster.
In addition, the fraction of the volume of the main
cluster in which turbulence is injected (the volume swept by the 
infalling subclusters) depends only on 
the mass ratio of the two merging clusters, provided that the distribution
of the accreted mass--fraction does not strongly depend on the cluster
mass (Lacey \& Cole 1993).
The combination of this two items yields a self--similarity
in the injection of turbulence in the ICM:
the energy of such turbulence should scale with the cluster
thermal energy and the turbulent energy should scale with
virial mass with a slope $D_{sem} \sim 1.67$ (C\&B05).

\begin{figure}
\includegraphics[width=0.49\textwidth]{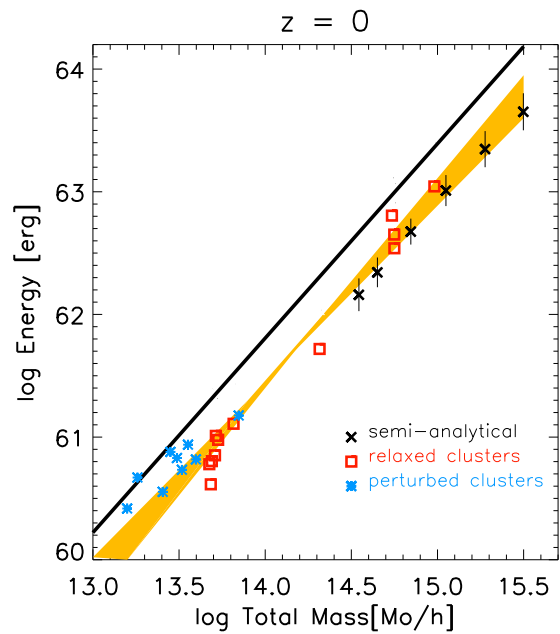}
\caption{Comparison between the thermal and turbulent scaling at zero redshift, 
for 12 ``relaxed'' (i.e. $\xi < 0.5$) galaxy clusters, 9 ``perturbed'' (i.e. $\xi \geq 0.5$)
 clusters and semi-analytical average data with $1 \sigma$ errors. The black line shows
 the thermal scaling of the whole simulated sample, while the
orange band encloses, within $1 \sigma$ errors, the scaling of the ``relaxed'' sample alone and
the scaling with the 9 ``perturbed'' object added.}
\label{fig:ross}
\end{figure}

In Fig.\ref{fig:ross} we report the integral of the turbulent energy
(injected in the ICM up to the present time) versus the
cluster mass, as estimated under the C\&B05 approach with 360
merging trees of massive galaxy clusters,  
together with the measures done on our hydrodynamically
simulated clusters: the two scalings are consistent within $1 \sigma$ errors.
The two approaches are complementary, since semi--analytical calculations 
can follow the properties of $>10^{15}$M$_{\odot}$
clusters which are rare in numerical simulations due to the limited
simulated cosmic volume, and so provide a physical extrapolation
of the trend derived by our numerical simulations.
This strengthens our claim that the turbulent velocity fields detected
in simulated clusters are actually real turbulent fields
supplied by the mass accretion process acting in galaxy clusters. 

Both estimates of the turbulent energy in galaxy clusters
show an overall content of turbulence which ranges from 25\% to 35\% of the thermal
one, in the  $(R_{\rm{vir}})^{3}$ region. This should be considered
as an upper limit of the turbulent energy content at a given time,
because simulations do not contain appropriate recipes for the 
dissipation of the turbulent eddies at the smallest scales (for the sake of 
comparison, the semi--analytical calculations in Fig.\ref{fig:ross} were
conceived to focus on the injection of turbulence in the whole cluster life).

Fig.\ref{fig:ross} highlights the different behaviour of ``perturbed'' (i.e. $\xi \geq 0.5$)
and ``relaxed'' clusters in the turbulent energy -- mass plane.
As discussed in Section \ref{sec:laws} the presence
of ``perturbed'' clusters/groups introduces a bias in the properties of 
the overall simulated cluster population. In this case the complete sample
of our simulations would be more representative of rich environments and superclusters,
with the smaller structures beeing more perturbed (and turbulent) than those
in other environments. This would cause a systematic flattening of the 
measured turbulent energy vs mass correlation.

\section{Conclusions}
\label{sec:conclusion}
Our sample of simulated galaxy clusters shows a well--constrained
power--law scaling between the kinetic energy of ICM turbulent motions
and the cluster mass. The slope of the scaling does not
show any evident dependence on the resolution adopted to detect the
pattern of turbulent motions, while the normalization shows a slight
dependence on resolution. Rather independently of the cluster mass, the turbulence
injected during the cluster life has an energy budget of the order of
25\% --35\% of the thermal energy at redshift zero.
If small perturbed clusters are included in the sample this affects 
the scaling law
of the turbulent energy making it slightly flatter than the thermal case;
we expect that this should be the case in supercluster regions.
Our scaling is in line with the semi--analytical findings of 
C\&B05: although they used a completely independent method, when plotted 
together the datapoint of the two approaches prove the scaling over more than
two decades in cluster mass. Semi--analytical calculations give a simple
physical explanation of the scaling laws in term of the ${\rm PdV}$ work
done by the infalling subclusters through the main ones, and strenghten
the physical nature of the turbulent field found in simulations.
The inclusion of cooling processes within our simulations 
is not expected to modify our conclusions,
because the average cooling time for the large cluster regions
considered here is longer than an Hubble time.
Cooling may play an important role in innermost regions,
where only a minor part of the turbulent energy is stored,
however the inclusion of cooling in simulations would also
require the implementation of feedback mechanisms -- like galactic winds and
bubble inflation by AGNs -- in order to prevent un--physical massive
cooling flows.
These processes should also 
introduce additional turbulence 
and further studies are required to
understand how such processes might effect the reported correlations.

 \section*{acknowledgements}
  F. V. thanks Riccardo Brunino, Carlo Giocoli and Marco Montalto
 for very useful discussions. 
The simulations were carried out on the IBM-SP4 machine at the
``Centro Interuniversitario del Nord-Est per il Calcolo
Elettronico'' (CINECA, Bologna), with CPU time assigned under an
INAF-CINECA grant, on the IBM-SP3 at the Italian Centre of
Excellence ``Science and Applications of Advanced Computational
Paradigms'', Padova and on the IBM-SP4 machine at the
``Rechenzentrum der Max-Planck-Gesellschaft'' at the
``Max-Planck-Institut f\"ur Plasmaphysik'' with CPU time assigned
to the ``Max-Planck-Institut f\"ur Astrophysik''. K.~D.~acknowledges
 support by a Marie Curie Fellowship of the
European Community program "Human Potential" under contract number
MCFI-2001-01227. R. C. \& G. B. acknowledge partial support from 
the MIUR undergrant PRIN2005.

\end{document}